\documentclass[letterpaper]{JHEP3}  
\usepackage{epsfig} 
\usepackage{graphicx} 
\def\r{\rho}

\title{On the Singularity Structure and Stability of Plane Waves}

\author{ 
Donald Marolf$^\dagger$ and 
Leopoldo A.  Pando Zayas$^{\ddagger}$ \\
$^\dagger$ Physics Department, Syracuse University, Syracuse, New York 
13244 USA\\ 
$^\ddagger$ 
Michigan Center for Theoretical  
Physics,
Randall Laboratory of Physics, The University of  
Michigan,
Ann Arbor, MI 48109-1120 USA\\
$^\ddagger$ 
School of Natural Sciences,
Institute for Advanced Study,
Princeton, NJ 08540}

\date{October, 2002} 
 
\abstract{
We describe various aspects of plane wave backgrounds. 
In particular, we make explicit a simple criterion for singularity 
by establishing a relation between Brinkmann metric entries and 
diffeomorphism-invariant curvature information. We also address  the 
stability of plane wave backgrounds by analyzing the fluctuations of generic
scalar modes. We focus our attention on cases where after fixing the
light-cone gauge the resulting world sheet fields appear to have
negative ``mass terms". We nevertheless argue that these backgrounds may
be stable. } 
\preprint{SU-GP-02/10-1, MCTP-02-52} 
 
\begin{document}

\section{Introduction and summary}
Plane wave spacetimes have special properties that motivate their
study in both General Relativity and string theory. 
Due to the presence of a covariantly  constant null Killing 
vector and the specific structure of the curvature invariants, it was 
suggested early on that $\alpha'$ corrections were under control for 
stringy plane wave backgrounds.  Along 
these lines  some concrete realizations were found in the context of 
WZW models providing examples of exact strings backgrounds 
on curved spacetimes \cite{old}
(see \cite{Trev} and references therein). More recently, 
renewed interest in plane wave 
backgrounds has resulted from the realization that
certain plane wave backgrounds are
maximally supersymmetric solutions of eleven dimensional 
and IIB supergravity \cite{joseM,joseIIB} (see also \cite{tratado}). 
More remarkable is the fact that some of these backgrounds allow 
for exact quantization in the light-cone gauge \cite{metsaev} 
(see also \cite{tsestring}), thus providing examples of tractable 
string theory backgrounds with Ramond-Ramond (RR) fluxes on 
curved spacetimes. Finally, 
based on the fact that 
some of the maximally supersymmetric plane waves 
can be obtained as Penrose-G\"uven  limits \cite{pen1,gueven} of $AdS_p\times S^q$ backgrounds 
\cite{josepen}, a 
gauge theory interpretation based on the AdS/CFT correspondence 
was proposed in \cite{BMN}.

These special properties have motivated the 
study of many generalizations.  One particularly prolific 
direction is the study of Penrose limits of known supergravity 
backgrounds.  The Penrose limit, generalized to include form fields and 
fermions by G\"uven, yields a plane wave metric. Some of these 
metrics are known, however many present new and unexpected features. 
This situation motivated us to study a 
fairly general class of such backgrounds. 

A particular class of interest are metrics which in the string frame
have the   so-called
Brinkmann \cite{Brinkmann}   form
 
\begin{equation} 
\label{metric} 
ds^2 = -4dx^+ dx^- - (x^i \mu^2_{ij}x^j) (dx^+)^2 + \delta_{ij} dx^i dx^j, 
\end{equation} 
where the matrix $\mu^2_{ij}$ is an arbitrary function of  $x^+$.
Any other fields should depend only on the coordinate $x^+$.
Furthermore,   the field strengths of  any gauge fields should
annihilate the vector field  $\frac{\partial}{\partial  x^-}$.  It is
clear  that the $x^-$ translation $x^- \rightarrow x^- + \epsilon$
with all other coordinates held constant  is a symmetry,  so that
$\frac{\partial}{\partial x^-}$ is a (null) Killing field.

We are interested in two very natural questions that arise in the 
context of Penrose limits of string theory backgrounds: (i) When is a 
plane wave metric singular? (ii) When is a plane wave background stable?
Since scalar curvature invariants vanish for plane waves, it is
useful to identify diffeomorphism invariant characterizations of
the properties of (\ref{metric}).
The question 
of stability appears when some of the eigenvalues of 
the matrix $\mu^2_{ij}$ are negative. Some comments on these backgrounds
have already  appeared in \cite{GPZS,HS1,HS2}.   The case where $\mu^2_{ij}$
has at least one negative eigenvalue is interesting because 
this seems to imply the existence 
of worldsheet tachyonic fields. 
Physically, such negative eigenvalues of $\mu_{ij}^2$
imply that objects are
stretched and separated by gravitational tidal forces.  The effect is
similar to that felt in de Sitter space, though without the formation
of horizons.

In section \ref{sing} we make explicit a criterion for 
detecting curvature singularities in plane wave backgrounds by explicitly relating a 
diffeomorphism invariant object to the structure of the metric.  Our 
criterion provides an immediate answer to the question of the 
singularity structure.

The question of stability is most cleanly formulated in
time-independent settings.  For this reason, we focus mainly on the
case where $\mu_{ij}^2$ is independent of $x^+$.  
In section 3, we consider a scalar mode as a typical example of a perturbation.
We study both the classical dynamics and 
quantization in plane
wave backgrounds.  We refer the reader to \cite{Dom} for a
stability analysis of other supergravity fields. We then proceed to stringy 
issues in section \ref{strings}. In the discussion section we point out
a number of open issues concerning string interactions. 
Our work has a certain overlap with that of \cite{Dom}, though the results
were obtained independently and simultaneously.

%%%%%%%%%%%%%%%%%%%%%%%%%%%%%%%%%%%%%%%%%%%%%
\section{Plane waves: Singular or Regular}
\label{sing}

In order to arrive at a criterion for regularity 
we need to describe some aspects of the geometry of plane waves.  We
proceed quickly since most of
these results are at least implicitly discussed in the literature (especially in
the classic works   \cite{Brinkmann,BPR,JH,AP,EK,exact}).
Nevertheless,  it is
useful to have such results in explicit form.

\subsection{Symmetries}
\label{sym}

Any metric of the Brinkmann form (\ref{metric}) has a rather high degree
of symmetry.  Killing's  equation is:
\begin{equation} 
0 = \xi_{\mu ; \nu } + \xi_{\nu; \mu}= \partial_\nu \xi_\mu +
\partial_\mu \xi_\nu -2 \Gamma_{\nu \mu}^\alpha \xi _\alpha,
\end{equation} 
and the only nonzero Christoffel symbols (with all indices down) are
\begin{eqnarray} 
\Gamma_{++i} &=& -\mu^2_{ij} x^j = - \Gamma_{i++}, \cr  \Gamma_{+++}
&=& -\frac{1}{2} x^i \bigl(  \frac{\partial}{\partial x^+} \mu^2_{ij}
\bigr) x^j,
\end{eqnarray} 
We use the conventions of Misner, Thorne, and Wheeler  \cite{MTW}.

Consider a Killing field with $\xi_-=0$ and with $\xi_i$ a function
only of $x^+$.   A lengthy but straightforward calculation shows that
all Killing fields in fact satisfy the first of these conditions while
the second condition holds for all Killing fields except for possible
rotational symmetries of $\mu_{ij}$.  Given these conditions,
Killing's equation reduces to
\begin{eqnarray} 
\label{first} 
&\partial_+ \xi_+ = x^i \mu^2_{ij} \xi^j,& \ \ \ \rm{and} \cr
&\partial_+ \xi_i  + \partial_i \xi_+  =0.&
\end{eqnarray} 
Latin indices are freely raised and lowered with  the flat Euclidean
metric $\delta_{ij}$.  Combining the above two equations yields
\begin{equation} 
\label{second} 
\partial_+^2 \xi_i  + \mu^2_{ij} \xi^j  =0,
\end{equation} 
which has solutions for all initial data  $\xi^i(0)$, $\partial_+
\xi^i(0)$.   The general solution for $\xi_+$ is then given by
\begin{equation} \xi_+ = - x^i  \partial_+ \xi_i + c ,  \end{equation} 
where $c$ is an arbitrary constant.  Note that we have identified
enough Killing fields to show that each null surface $x^+=constant$ is
a homogeneous surface.

The detailed properties of the Killing fields depend on the choice of
initial data.  For definiteness, let us consider the time independent
case, where $\partial_+$ is also a Killing field.  Then the $\xi^i$
form a collection of harmonic or inverted harmonic oscillators. Some
choices of initial data thus lead to `corkscrew' motions which wind or
rotate around the origin.  Such Killing fields are no doubt related to
the  rotating D-branes of \cite{ST}.

\subsection{Geodesics and Curvatures} 
\label{GandC}
 
Let us now consider the curve $x^i=0, x^-=0$.  Since the metric
(\ref{metric}) is symmetric under the reflection   $(x^i,x^-, x^+)
\rightarrow (-x^i,-x^-,-x^+)$,   an acceleration  vector for this curve
at $x^+=0$ must be invariant under this symmetry.   
As a result, this vector can
point only along the curve.  But since the origin of $x^+$ is arbitrary,
we see that the curve is a geodesic.
Direct calculation then shows  that $x^+$ is an associated affine
parameter.

A large family of null geodesics  can be obtained by applying
symmetries to the curve $x^i=0,x^-=0;$ i.e., by acting with the
translation $e^\xi$.  These geodesics satisfy $x^+ = \lambda$, $x^- =
\frac{1}{4} (\xi^i \partial_+ \xi_i) - c$, and $x^i(x^+) = \xi^i(x^+)$ with $\lambda$ an
affine parameter and $\xi$ some Killing field of the form discussed in
section \ref{sym}.  The fact that these curves  trace out geodesics explains
why equation (\ref{second}) will  look familiar to readers who have
not previously studied the  Killing symmetries of these spacetimes but
who have studied  geodesic motion.  This
family contains all of the null geodesics except those that follow the
orbits of $\frac{\partial}{\partial x^-}$.

Suppose that   two geodesics are related by a flow   through a unit
Killing parameter  along some Killing field ${\bf \xi}$ of the form
described in section \ref{sym}.  Then the distance separating these
geodesics along any surface $x^+ =constant$  is given by $|{\bf \xi}|
= \sqrt{|\sum_i \xi^i \xi^i|^2}.$    Since $x^+$ is an affine
parameter for this family of geodesics, it  is also meaningful to
speak of the relative velocity $\partial_+ \xi^\alpha$  and
acceleration $\partial_+^2 \xi^\beta$ of these geodesics, so long as
we keep in mind the freedom to rescale any affine parameter by an
overall  (i.e., $x^+$ independent) constant and to shift its origin.  
In particular, this information
is diffeomorphsim invariant modulo the potential scaling and shift.  
Note that these relative
accelerations   measure gravitational tidal forces and, if the
geodesics are infinitesimally separated,  they are given by  certain
components of the Riemann tensor:
 
\begin{equation} 
\partial^2_+ \xi^\nu = - R_{\alpha \beta \gamma}^\nu \xi^\beta
t^\alpha t^\gamma,
\end{equation} 
where $t^\alpha$ is the tangent vector to the geodesic.    If we take
one of the geodesics to be at $x^i=0$, then this tangent  vector is
just $t^\alpha \partial_\alpha = \frac{\partial}{\partial x^+}$.
 
Thus for Killing vector fields $\xi$ and $\xi'$ we have
\begin{equation} 
\label{curv} 
\xi^i \mu_{ij}^2 \xi'{}^j = - \xi_i \partial_+^2 \xi'{}^i  = \xi_\nu
R_{\alpha \beta \gamma}^\nu \xi'{}^\beta  t^\alpha t^\gamma,
\end{equation}   
so that the matrix-valued function $\mu^2_{ij}(x^+) = R_{i+j+}(x^+)$
describes diffeomorphism-invariant curvature information (up to an
overall scale and shift of the origin of $x^+$).  In particular,  the
divergence of any component of $\mu^2_{ij}$  at some value of $x^+$
represents a singularity where invariantly defined  components of the
curvature tensor   diverge\footnote{The diverging tidal forces  are
sufficient to overcome any internal  forces within an object
approaching this singularity.  This  effect causes a diverging
relative acceleration between two parts  of some object.  However, the
actual distortion of the object  (e.g., the change in relative
position of these parts) is a second  integral of the acceleration and
may in some cases remain finite.  As s result, one sometimes makes the
(diffeomorphism invariant)  distinction between `strong' and `weak'
singularities \cite{weak} based on whether   an object approaching
this singularity is infinitely   distorted or is distorted only by a
finite amount.  Which occurs here  depends on whether ${\bf \xi}$
itself diverges,   which in turn is determined by the rate at which
$\mu^2$ diverges.}.

%%%%%%%%%%%%%%%%%%%%%%%%%%%%%%%%%%%%%%%%%%%%%%%%%%%%%%%%%%%%%%%%%%%%%%%%%%%
\subsection{Penrose Limit of Schwarzschild Black Hole in AdS}
%%%%%%%%%%%%%%%%%%%%%%%%%%%%%%%%%%%%%%%%%%%%%%%%%%%%%%%%%%%%%%%%%%%%%%%%%%%
Taking the Penrose limit of supergravity backgrounds dual to gauge theories
with a mass scale has been one of the most natural generalizations of 
the BMN construction. In most cases intuition from the field theory side
has helped in clarifying the effect of the Penrose limit on the original
background. However, a universal understanding of the Penrose limit in intrinsically 
gravitational terms is lacking in the recent
literature. Here we elaborate on a concrete
example\footnote{The Penrose limit 
of this background was formally considered in  \cite{Cobi}.} -- the
Schwarzschild 
black hole in AdS. Our intention 
is to provide a 
careful
interpretation of the effect of the limit on
the original background along the lines 
of the work by Geroch \cite{geroch}. This approach was implicitly 
assumed also in \cite{tratado}.

The metric we consider is

\begin{equation}
\label{bh}
ds^2=-({r^2\over R^2}+1-{M\over r^2})dt^2 + ({r^2\over R^2}+1-{M\over
r^2})^{-1}dr^2 + r^2 d\Omega_3^2 
+ R^2\bigg[d\psi^2 + \sin^2\psi
d\Omega_4^2\bigg].
\end{equation}
The above metric describes the Schwarzschild black hole in AdS in global 
coordinates. 
To perform the Penrose limit we 
 consider a null geodesic in the directions determined by
$(t,r,\psi)$. The effective Lagrangian that is
equivalent to solving the geodesic equations is:
\begin{equation}
\label{bhgeodesic}
{\cal L}=-({r^2\over R^2}+1-{M\over r^2})\dot{t}^2 + ({r^2\over R^2}+1-{M\over
r^2})^{-1}\dot{r}^2+ R^2 \dot{\psi}^2,
\end{equation}
where dot represents derivative with respect to the affine parameter $U$. 
The aim is, following Penrose's prescription \cite{pen1}(see also
\cite{tratado}), 
to find new coordinates
$(U,V,Y^i)$ where the metric takes the form 
\begin{equation}
ds^2=dV(2dU+\alpha dV + \beta_i dY^i) + C_{ij} dY^i dY^j,
\end{equation}
where $\alpha, \beta_i$ and $C_{ij}$ depend on all the the coordinates.
Once the metric is in this form the Penrose limit is given by the scalings
\begin{equation}
\label{pl}
ds^2=\Omega^{-2}ds^2(\Omega), \quad U=u, \quad  V=\Omega^2 v, \quad
Y^i=\Omega y^i, \quad \Omega\to 0. 
\end{equation}
Note that $\partial_U$ is a null vector  since 
\begin{equation}
g_{UU}=G_{tt}\dot{t}^2+G_{rr}\dot{r}^2 + G_{\psi \psi}\dot{\psi}^2=0
\end{equation}
by definition of a null geodesic. The equations of motion following from
(\ref{bhgeodesic}) are particularly simple since the effective Lagrangian 
 does not
depend explicitly on $t$ or $\psi$:
\begin{equation}
\label{U}
\dot{\psi}={\mu\over R^2}, \quad \dot{t}={E\over {r^2\over R^2}+1-{M\over
r^2}}, \quad \dot{r}^2+{\mu^2 \over R^4}r^2 -{\mu^2 \, M\over R^2 \,
r^2}=E^2-{\mu^2\over R^2}.
\end{equation}
These three equations completely determine the dependence of the
coordinates $(r,t,\psi)$ on the affine parameter $U$. To complete the
coordinate transformation we use the conditions that $G_{UV}=1$ and 
$G_{U\Phi}=0$. The final
coordinate change is the of the form
\begin{equation}
r=r_U, \quad t=t_U -{V\over E}+\mu \Phi, \quad \psi={\mu\over R^2} U + E\Phi,
\end{equation}
where $r_U$ and $t_U$ are functions only of $U$ determined by
(\ref{U}). Taking the Penrose limit
following (\ref{pl}) we find the following metric in Rosen coordinates
\begin{equation}
\label{brosen}
ds^2=2dudv +(E^2R^2-\mu^2
f(r))d\phi^2 + r^2 ds^2({\bf R}^3) + R^2 \sin^2({\mu\over R^2}U) ds^2({\bf R}^4),
\end{equation}
where $f(r)={r^2\over R^2}+1-{M\over r^2}$ and $r$ is related to $u$ as
\begin{equation}
\label{ru}
2{\mu\over R^2}u=\arctan{\mu r^2 /R^2 -E^2R^2/2\mu\over 
\sqrt{(E^2-{\mu^2\over R^2})r^2-{\mu^2\over R^4} r^4+{\mu^2 M\over R^2}}}.
\end{equation}
An arbitrary metric in Rosen coordinates 
\begin{equation}
ds^2=2du dv + C_{ij}(u) dy^i dy^j,
\end{equation}
can be presented in Brinkmann form 
\begin{equation}
ds^2=-4dx^+dx^- + \bigg[\sum\limits_{i,j}A_{ij}(x^+)x^i x^j\bigg] (dx^+)^2
+ \sum\limits_{i}dx^idx^i,
\end{equation}
by means of the following coordinate transformation:

\begin{equation}
u=2x^+, \quad v= x^- -{1\over 2}\sum\limits_{i,j}M_{ij}(x^+)x^ix^j, \quad 
y^i=\sum\limits_i Q^i_j(x^+)x^j
\end{equation}
where 
\begin{equation}
C_{ij}Q^i_kQ^j_l=\delta_{kl}, 
\quad C_{ij}(Q'{}^i_jQ^j_l-Q'{}^i_kQ^j_l)=0, \quad M_{ij}=C_{kl}Q'{}^k_lQ^l_j,
\end{equation}
where prime indicates derivative with respect to $x^+$. 
In the particular case when $C_{ij}=a^2_i(x^+)\delta_{ij}$ is diagonal 
we have that 
\begin{equation}
A_{ij} = {(a_i(x^+))''\over a_i(x^+)}\delta_{ij}
\end{equation}
For the Schwarzschild black 
hole we obtain a metric that
can be written in Brinkmann coordinates as:
\begin{equation}
ds^2=-4dx^+dx^- -{\mu^2\over R^4}\bigg[\sum\limits_{a=1}^4x_a^2 
+(1+{MR^2\over r^4})\sum\limits_{b=5}^7 x_b^2 + (1-3{MR^2\over r^4})x_8^2\bigg]
(dx^+)^2+ \sum\limits_{i=1}^8dx_i^2
\end{equation}

One question that naturally arises when considering Penrose 
limits of metrics containing dimensionful parameters is what happens
to these parameters under the rescaling. In what follows we address 
this question in the specific case of  AdS-Schwarzschild but our approach is 
universal. 
First, note that the Penrose limit can be thought of in two parts (performed
simultaneously).  One is an overall scaling of the metric by $\Omega^{-2}$.
This takes the original spacetime $ds^2$ to a physically different spacetime
$ds^2_\Omega = \Omega^{-2} ds^2$.  In the case at hand, it takes
us from one AdS-Schwarzschild metric to another.
Now, the cosmological constant and black hole mass
are diffeomorphism-invariant quantities.  Suppose we calculate
the cosmological constant and black hole mass from $ds^2_\Omega$ and
use these to define the ($\Omega$-dependent) quantities $R_\Omega$
and $M_\Omega$.  Then dimensional analysis is enough to tell us that
we have
 
\begin{equation}
R_\Omega = R/\Omega
\end{equation}
 
and
 
\begin{equation}
M_\Omega = M/\Omega^2.
\end{equation}
 
Note that these two parts are interpreted differently in the dual
field theory.  The rescaling of the metric and corresponding change
from one spacetime to another requires that we alter the dual field
theory to match.  In particular, we must take
$g_{YM}N$ in the field theory to become infinitely large
in a way that matches $R_\Omega$.  
This is the content of the statement ``$R = \Omega^{-1}$'' commonly
associated with this limit \cite{BMN}.
In particular, one does not rescale 
$R$ in addition to taking the Penrose limit; rather, the
rescaling of $R$ is a consequence of interpreting the strict Penrose
limit in the dual field theory. Tracing through the duality map
should also tell us on the field theory side how the temperature of the
thermal bath corresponding to the black hole varies with $\Omega$. Not
surprisingly, we find that the temperature of the black hole is
$T_\Omega=\Omega T$ and therefore goes to zero in the limit as is compatible
with our knowledge that the ``black hole" disappears (see
\cite{HR} for a discussion of the possibility of black holes in plane waves). 

In much the same way that we defined $M_\Omega$ and $R_\Omega$
above, one may define $E_\Omega$ and $\mu_\Omega$ by 
calculating them from (\ref{U}) applied to the favored null 
geodesic in the spacetime $ds^2_\Omega$. The key point is that 
the geodesic remains
the same in terms of the coordinates $\tilde t = t/R,\tilde r = r/R,$ 
and $\psi$ and that the parametrization remains unchanged (we always
use affine parameter $U$).  (The geodesic also remains unchanged in terms of 
$t,r, \psi$, but $\tilde t, \tilde r,$ and $\psi$ are nicer as
the rescaled metric maintains the same form except for the replacement
of $R$ with $R_\Omega$.)  Thus, $\dot{\psi} = \mu/R^2$ is independent
of $\Omega$ and $\dot{\tilde t} = E/f(r)R$ is independent of $\Omega$.
In other words, we may identify

\begin{equation}
E_\Omega = E/\Omega
\end{equation}
and

\begin{equation}
\mu_\Omega = \Omega^{-2} \mu.
\end{equation}
It is natural to assert that such scalings
are the supergravity dual of the statement on 
the field theory side that we are focusing  
on states with large energy and large R-charge.

To further simplify the above metric
we introduce $q = ER/\mu$ and 
$\tilde r = r/(qR)$ and $\tilde m = M/(qR^2)$. Then we have
\begin{equation}
\label{BHpw}
ds^2=-4dx^+dx^- -
\bigg[\sum\limits_{a=1}^4x_a^2
 +(1+{\tilde m\over {\tilde r}^4})\sum\limits_{b=5}^7 x_b^2 + (1-3{\tilde 
m\over {\tilde r}^4})x_8^2\bigg](d{{\mu x^+} \over R^2} )^2
+ \sum\limits_{i=1}^8dx_i^2.
\end{equation}
Furthermore, note that the equation (\ref{ru}) relating $u$ and $r$ can
be written:

\begin{equation}
{{\mu x^+} \over R^2}  = \arctan {{\tilde r^2 - 1/2}  \over {\tilde r^2 
- \tilde r^4 + \tilde m} },
\end{equation}
so that the entire final metric is specified by the parameter $\tilde m$.
It is worth noting that the same background is obtained as the limit of 
non-extremal D3-branes in Poincare coordinates. Now we have an answer to
the question of what happened to the $r=0$ singularity of the original
black hole. According to the criterion above, a singularity 
remains at $r=0$ in the Penrose limit.

%%%%%%%%%%%%%%%%%%%%%%%%%%%%%%%%%%%%%%%%%%%%%%%%%%%%%%%%%%%%%%%%%
\section{Field theory on plane waves with negative eigenvalues} 
\label{FT}

One of the intriguing features of the background (\ref{BHpw}) is the
negative eigenvalue (associated with $x_8$) of the matrix
 $\mu^2_{ij}$ that arises for small $r$. If one attempts to quantize string
theory in the light-cone gauge on the  background (\ref{BHpw}) one would
have a negative ``mass term" for the worldsheet boson $x_8$. One might
wonder whether this is a tachyonic instability. The appearance of these
negative ``mass terms" is commonplace when studying Penrose limits of
backgrounds with a mass scale \cite{GPZS,roy}.

Here we investigate plane wave backgrounds in the field theory
approximation.  To be specific, we assume that the spacetime
(\ref{metric})  is a classical solution to some gravitating theory and
consider small perturbations about this background.   For
definiteness, we assume that the perturbations can be modeled by
massless scalars.   Below, we consider the case where the matrix
$\mu^2_{ij}$ has at least one negative eigenvalue.  The weak energy
condition then implies that $\mu^2_{ij}$ also has at least one positive
eigenvalue.

Let us take a moment to put our problem into perspective.  Our real
concern is the  stability of the spacetime.  This concern is raised by
the negative eigenvalues of $\mu^2_{ij}$, which for example appear as
negative mass-squared terms in the light-cone string sigma-model.  One
might therefore think that our spacetime has tachyonic perturbations.
But such perturbations can be identified in a classical stability
analysis.  As a result, one would expect that the most important
features can be seen directly in a classical analysis of linear fields
propagating on the plane wave background.  One would also expect such
perturbations to be qualitatively similar for all massless fields.
Indeed, while we will return to quantum effects in section
\ref{freeQFT} and to string effects in section \ref{strings},  in the
end we will see that the most important features are captured by the
classical  massless scalar field which we investigate below.
 
\subsection{Massless classical field theory on the plane-wave spacetime} 
\label{classical}

Consider a massless scalar field minimally coupled to the string
metric.  Such a field satisfies the equation of motion
 
\begin{equation} 
\label{FreeEOM} 
\nabla^2 \phi =  0,
\end{equation} 
where $\nabla^2$ is the Laplace-Beltrami operator  associated with the
string metric.  Recall that the curvature scalar $R$ and the squared
field strength $F^2$ of any gauge field both vanish in our
background. Thus, dilaton perturbations  also satisfy (\ref{FreeEOM}),
as do scalar fields with arbitrary curvature couplings.  Perturbations
of other supergravity fields are discussed in \cite{Dom}, where it is
shown that their equations of motion take a similar form in an
appropriate gauge.
 
For the metric (\ref{metric}), the wave equation (\ref{FreeEOM}) 
is proportional to the condition
\begin{equation} 
\label{EOM} 
[- 4 \partial_- \partial_+ + (x^i \mu^2_{ij} x^j) \partial_-^2  + 4
\delta^{ij} \partial_i  \partial_j ]\phi =0.
\end{equation}
It is convenient to expand $\phi$ in a basis of modes.  Let us
consider first the case where $\mu^2_{ij}$ is independent of $x^+$, so
that it makes sense to choose modes $\phi_{p_-,p_+,\beta}$  which
satisfy
\begin{eqnarray} 
\partial_- \phi_{p_-,p_+,\beta} &=& i p_- \phi_{p_-,p_+,\beta}, \ \ \
{\rm and} \cr \partial_+ \phi_{p_-,p_+,\beta} &=& i p_+ \phi_{p_-,
p_+,\beta}.
\end{eqnarray} 
Here, $\beta$ is any index that  labels the remaining degeneracy  of
the modes.  If desired, explicit solutions are readily obtained
in terms of Hermite polynomials.

The central question is whether such perturbations are stable.  In
other words, we must ask whether there can be a complete set of modes
$\phi_{p_-, p_+,\beta}$ with real $p_+,p_-$.  The fact that the `mass
term' $(x^i \mu^2_{ij} x^j)$ in (\ref{EOM}) can be negative suggests
that complex frequencies may be required.  Thus, it is worth
investigating the matter in some detail.

Let us pause for a moment to be sure that we understand what is meant
by a complete set of modes.  Clearly, the fundamental requirement is
that the modes be sufficient to analyze any reasonable set of initial
data.  Initial data is typically imposed on a spacelike Cauchy
surface, but even the  identification of a complete Cauchy surface is
nontrivial in this spacetime\footnote{While this spacetime is not
globally hyperbolic \cite{Penrose}, as one would
do in AdS space, it presumably makes sense to speak of Cauchy surfaces
once boundary conditions are specified at infinity. However, even in
this sense Cauchy surfaces are nontrivial to identify explicitly.}.

The situation is much improved by imposing an infrared regulator in
the $x^i$ directions.  We therefore impose boundary conditions which
restrict $x^2 < L^2$.  Note that such boundary conditions cannot
affect local dynamics in small regions of spacetime
near the center of the plane wave.

However, no matter how large $L$ may be, signals will propagate
from the boundary to the origin in an interval of $\Delta x^+$
no larger than $\pi/\mu_1$, where $\mu_1^2$ is the largest eigenvalue
of $\mu^2_{ij}$ (see e.g. \cite{MR} for an analysis of light-cones
in plane wave spacetimes).  The issue of decoupling of the boundary condition is
therefore subtle as $ L \rightarrow \infty$.  

Let us take guidance from the simple BFHP case $\mu^2_{ij} = \mu^2 \delta_{ij}$
with $\mu^2 > 0$.  In particular, since the Ricci scalar vanishes,
massless scalars with any curvature coupling have the same dynamics.
On the other hand, it is known \cite{BN} that the BFHP spacetime can
be conformally embedded in the Einstein static universe ($S^9 \times R$),
where it covers all but a single null line which forms the boundary.
For the conformally coupled case, it is therefore clear that imposing
an appropriate a boundary condition at $x^2=L^2$ will reduce
to using `transparent' boundary conditions on this null line in the limit
$L \rightarrow \infty$; i.e., the end result is a dynamics defined
directly on the Einstein static universe without giving any special status
to the null line which forms the conformal boundary of the BFHP spacetime.

We therefore assume that the case of general $\mu^2_{ij}$ behaves similarly
and that there is a natural set of boundary conditions which may be
imposed at $x^2 =L^2$ for large $L$.
Although no conformal compactification is known for this general case and
one might expect the asymptotic structure to change when an eigenvalue
of $\mu^2_{ij}$ becomes zero or negative,  
we are encouraged to make this assumption by the results of \cite{MR}.
In particular, it was shown there that the causal boundary in the
sense of \cite{GKP} of the 
general time-independent plane wave is identical to that of the BFHP case, 
consisting of a single null line.

Let us now choose some $\epsilon \ll L^2 |\tilde \mu^2|$, where $\tilde \mu^2$
is the
largest negative eigenvalue of $\mu^2_{ij}$.  The surface $x^+
= \epsilon x^-$ is spacelike in the region  $x^2 < L^2$ and, given the
boundary condition at $x^2 = L^2$, forms a Cauchy surface $\Sigma_\epsilon$
for this
region.  It is convenient to introduce a coordinate $z = \epsilon x^+
+  x^-$ along the surface and a coordinate $t = x^+ - \epsilon x^-$
transverse to the surface.  The associated momenta are $p_z =
\epsilon (1+ \epsilon^2)^{-1} p_+ + (1+ \epsilon^2)^{-1} p_-$  and
$p_t = (1+ \epsilon^2)^{-1} p_+ - \epsilon (1+ \epsilon^2)^{-1} p_-$.

The most transparent analysis would be to consider modes with
arbitrary real $p_z$ and which correspond to a complete set of modes
in the $x^i$ directions.  One would like to see if such modes lead to
real frequencies $p_t$.  However, the form of the equation of motion
(\ref{EOM}) makes this technically difficult.

As a result, we pursue a different strategy here.  While we will
neither prove or disprove stability, we do show that an infinite set
of stable modes exists for every value of $p_z$.  Furthermore, this
set includes modes with long wavelengths in the $x^i$ directions.  In
particular, many modes corresponding to negative eigenvalues of the
operator $H_{p_-} = - (x^i \mu^2_{ij} x^j) p_-^2  + 4  \delta^{ij}
\partial_i \partial_j$ are stable. Note that these  are just the
modes which originally motivated our search for instabilities.

\subsection{An argument for stability}
\label{arg}

We begin by restricting to modes with real $p_x$ and $p_t$, so that
$p_-$ is also real.  Though we have now assumed stability, the point
as outlined above will be to demonstrate that the class of stable
modes is extremely inclusive.  In particular, we will show that the
most obvious candidates for instabilities are in fact stable.

Now, for each real $p_-$, the operator $H_{p_-} = - (x^i \mu^2_{ij}
x^j) p_-^2  + 4  \delta^{ij} \partial_i  \partial_j  $ is
self-adjoint and its eigenstates yield a complete set of modes in the
$x^i$ directions. It is useful to think of this operator as the
Hamiltonian of a system of harmonic and inverted harmonic oscillators
with Dirichlet boundary conditions imposed at $x^2 = L^2$.  The
boundary conditions guarantee that the spectrum of $H_{p_-}$ will be
discrete and that it is meaningful to follow the $n$th eigenstate as a
function of $p_-$.  Thus, we may introduce the eigenvalue $E_n(p_-)$
corresponding to this state, and we may use $n$ as the additional
label $\beta$ introduced above\footnote{In non-generic cases  a
further discrete label will be required to remove a remaining
degeneracy.}.

The label $p_+$ becomes redundant as a mode $\phi_{p_-,p_+,n}$
satisfies
\begin{equation} 
\label{disp} 
[4 p_- p_+ + E_n(p_-) ]\phi_{p-,p+_, n} =0.
\end{equation}
Note that  if $E_n$ did not depend on $p_-$, this contribution  would
act like a two-dimensional mass term in (\ref{disp}).  The dependence
on $p_-$ means that we have instead arrived at a non-standard dispersion
relation whose analysis will require more care.

Let us therefore estimate $E_n(p_-)$ in various regimes.  
We will assume that the boundary conditions do not make a large
contribution. For small
$|p_-|$, the action of  $H_{p_-}$ on the $n$th eigenstate is dominated
by the kinetic term, $4  \delta^{ij} \partial_i \partial_j $.  Given
the infrared cutoff $L$, we must have
\begin{equation}
E_n \sim \frac{\pi^2 n^2}{4  L^2} > 0,  \ \ \ \ {\rm for} \ \ p_-^2
|\mu^2| L^4 \ll n^2.
\end{equation}
On the other hand, for fixed $n, \mu^2, L$ and large $p_-$ the
potential term will dominate and we will find
\begin{equation}
\label{lgp}
E_n \sim  p_-^2  \tilde \mu^2 L^2,  \ \ \ \ {\rm for} \ \ p_-^2
|\mu^2| L^4 \gg n^2,
\end{equation}
where $\tilde \mu^2$ is again the most negative eigenvalue of $\mu^2$.
Since $H_{p_-}$ depends smoothly on $p_-$, one expects $E_n(p_-)$ to
interpolate smoothly between these two regimes.  It will be important
in our analysis that $E_n$ is positive for small $p_-$, as small
momentum is the regime where tachyons are most dangerous.

Comparing these results with (\ref{disp}), we see that $p_+$ is linear
in $p_-$ for large $p_-$, but satisfies the dispersion relation for a
particle of {\it positive mass} for small $p_-$.  Graphs of dispersion
relations with these properties are sketched in figure \ref{dispplot} below
for large and small values of $L^2 \tilde \mu^2$.  These graphs show
explicitly that one may connect these two regimes in a manner such
that the tangent to the dispersion relation is always spacelike, so
that the group velocity of waves is less than one.  Most
fundamentally, it is clear that the dispersion relations provide real
values of $p_t$  for each $p_z$.  Thus, given $p_z$ and $n$, there are
unique associated real values of $p_-$ and $p_+$ for which the $n$th
eigenstate of $H_{p_-}$ defines a stable mode with frequency $p_+$.
\FIGURE{\epsfig{file=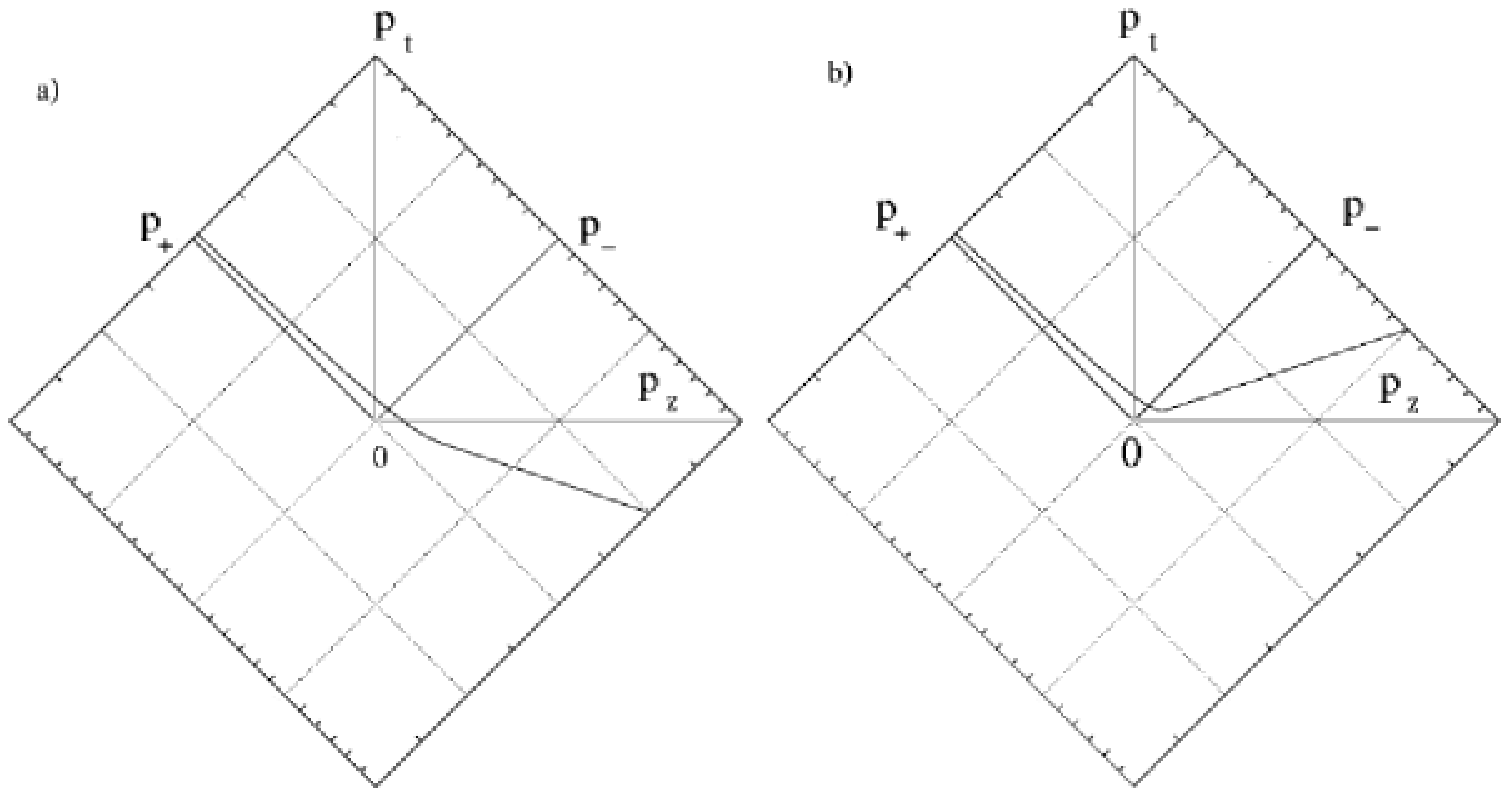}\caption{Plots of the $p_- > 0$
dispersion  relation for a) large $L \tilde \mu$ and b) small $L
\tilde \mu$.  Here $\epsilon =1$. The $p_- <0$ branches can be
obtained by inverting these diagrams through the
origin.}\label{dispplot}}

This verifies the claims made at the end of the previous
subsection.  Modulo our imprecision as to the boundary conditions, it
would constitute a proof of stability  if we could show
that the modes labeled by the pair $(p_z,n)$ formed a complete set of
states on $\Sigma_\epsilon$, but this is beyond the scope of our
work\footnote{Note that this is not equivalent to, and is in fact much
more subtle than, proving that these same modes form a complete set of
states on the surface $x^+=constant$; i.e., in $L^2(dx^-d^8x)$.  This
latter statement follows directly once the states are labeled by the
pair $(p_-,n)$, as this problem separates nicely into Fourier analysis
in $x^-$ and spectral analysis of $H_{p_-}$ }.

\subsection{Quantum field theory}
\label{freeQFT} 

It was argued above that the classical field is in fact stable.
Assuming that this is the case, we now show that no further
difficulties arise when the field is quantized.   In particular, we
will assume below that there is a complete set of states labeled by
real eigenvalues $p_-$.  Much
of this material has appeared before in \cite{Gibbons,HS1,HS2}, but 
we repeat it here for clarity and completeness. 

It is convenient to drop the requirement of translation
invariance in $x^+$ and to allow a general plane wave.  Thus, in this 
subsection we
require only that our field has an expansion of the form
\begin{equation} 
\phi =\int d p_- d \alpha \ a_{p_-,\alpha} \phi_{p_-,  \alpha},
\end{equation}
where $p_-$ ranges over $(-\infty, +\infty)$ and $a_{p_-, \alpha}$ is
the requisite operator valued coefficient for the mode $\phi_{p_-,
\alpha}$. Again, the mode $\phi_{p_-,  \alpha}$ is an eigenstate of
$\partial_-$ with eigenvalue $p_-$.

Some coefficients $a_{p_-, \alpha}$ will be creation operators and
some will be annihilation operators as determined by the commutation
relations.  As usual, these commutation relations may be computed from
the Klein-Gordon inner products of the corresponding  mode functions
(see, e.g., \cite{Bryce}):
\begin{equation} 
\label{CR} 
[a_{p_-,\alpha}, a_{  \tilde p_-\alpha}] = -i ( \phi_{p_-,\alpha},
\phi_{  \tilde p_-,\tilde \alpha})_{KG} = -i \int_\Sigma \phi_{p_-,
\alpha} \stackrel{\leftrightarrow}  {\partial}_n \phi_{\tilde
p_-,\tilde \alpha}  \sqrt{g_\Sigma} d\Sigma.
\end{equation} 
Our convention is that the Klein-Gordon inner product is linear  in
both entries and that the inner product of real functions is real.  In
(\ref{CR}), $n$ is the unit normal vector to the Cauchy surface
$\Sigma$  and $g_\Sigma$ is the determinant of the induced metric on
$\Sigma$.  Since the Klein-Gordon product is conserved, (\ref{CR}) may
be evaluated on any Cauchy  surface $\Sigma$.

It is well known that the treatment of plane wave backgrounds
simplifies in a light cone formalism.  This suggests that we take the
surface $\Sigma$ to be of the form $x^+ = constant$.  However, there
are several subtleties with this choice.  The first is that, as noted above, 
such
surfaces are not Cauchy surfaces.  However the
only causal curves which fail to cross such a surface are those that
asymptotically have constant $x^+$.  Such curves are a set of measure
zero.   In particular, each null geodesic crosses $\Sigma$ exactly
once unless it lies along an orbit of $\frac{\partial}{\partial x^-}$
(see, e.g.,  section \ref{GandC}).
 
The other two issues are of a more direct practical sort.  Since a
surface with $x^+ = constant$ is null, the normal cannot be normalized
and the induced measure vanishes.  However, these problems cancel
against each other as $d \Sigma \sqrt{g_\Sigma} \partial_n$ has a
well-defined limit when $\Sigma$ is approximated by a family of
spacelike surfaces $\Sigma_\lambda$.  We choose $\Sigma_\lambda$ to be
defined by $x^+ = f_\lambda(x^-,x^i)$ for, e.g.,
\begin{equation}
f_\lambda =  \frac{\tanh(x^-/\lambda)}{1 + \lambda^{-2} (x^i \mu^2_{ij}
x^j)^2}.
\end{equation}
It is not hard to show that $\Sigma_\lambda$ is spacelike for
sufficiently large $\lambda$, and that the family converges to the
surface $x^+=0$.

A straightforward calculation shows that in this limit we have
$\sqrt{g_\Sigma} \sim \lambda^{-1/2}$, while $\partial_n \sim
\lambda^{1/2} \partial_-$ and $d \Sigma \sqrt{g_\Sigma} \partial_n$
asymptotes to a constant times $dx^- d^8x \partial_-$.  As a result,
the Klein-Gordon inner product on $\Sigma$ takes the form
 
\begin{equation} 
\label{KGS} 
( \phi_{p_-,\alpha}, \phi_{  \tilde p_-,\tilde \alpha})_{KG} = A
\int_\Sigma \phi_{p_-,  \alpha} \stackrel{\leftrightarrow}
{\partial}_- \phi_{\tilde p_-,\tilde \alpha}  dx^- d^8x.
\end{equation} 
where $A$ is a positive normalization coefficient.
 
Thus, the Klein-Gordon inner product will be proportional  to $p_-$,
and $p_-$ will determine the all important sign of the commutation
relations.   Let us make the following standard choices for the mode
functions $\phi_{p_-, \alpha}$.  First, we take modes with opposite
$p_-$ and the same $\alpha$ to be complex conjugates, $\phi_{-p_-,
\alpha} = \overline{\phi_{p_-, \alpha}}$.  Second, we take modes with
distinct $\alpha$ to be orthogonal in $L^2({\bf R^8},
d^8x)$\footnote{Note  that orthogonality in $L^2({\bf R}^8, d^8x)$ is
preserved under evolution in $x^+$ since the light cone Hamiltonian
is self-adjoint in this space for each value of $p_-$.}.  Finally,
taking the normalization of our modes to be $1/\sqrt{A |p_-|}$ in
$L^2({\bf R^9}, d^8x dx^-)$, we find that the Klein-Gordon inner
product for modes  with $p_->0$ yields:
\begin{equation} 
\label{cr} 
[a_{p_-,\alpha}, a_{\tilde p_-\alpha}] = \rm{sign}(p_-) \delta(p_- +
\tilde p_-) \delta(\alpha, \tilde \alpha).
\end{equation} 
This is the characteristic algebra of creation and annihilation
operators.  Note that the reality of $p_-$ was critical in writing
$p_-/|p_-| = \rm{sign}(p_-)$.

{}From (\ref{cr}) we see that $a_{p_-,\alpha}$   is an annihilation
operator  for $ p_- >0$ and a creation operator for $p_- < 0.$  The
details of modes with $p_-=0$  exactly are a mere  matter of
convention since there are no normalizable $p_-=0$ modes.   Note that
when an eigenvalue of $\mu^2$  is negative, the spectrum of
$[4\delta^{ij} \partial_i    \partial_j  - (x^i \mu^2_{ij} x^j)
p_-^2]$ covers the entire real line so that    negative values of
$p_+$ can arise for any $p_-$ and, in particular, mode functions
associated with creation operators can have either sign of $p_+$.  In
this sense we will necessarily have negative energy particles.
However, we repeat that $p_+$ always  remains real so
that the modes of $\phi$ are stable.  The structure of the Fock  space
is much like that of a massless free field in Minkowski space.  In
particular, the vacuum is the unique normalizable state with $p_-=0$.
This rules out any pair creation of particles from the vacuum so that
the  vacuum is stable.  Since the theory is non-interacting, each
$n$-particle state is stable as well.

\subsection{Interactions in Quantum field theory} 
\label{intQFT} 
 
Suppose, however, that we now turn on some self-interaction for  our
massive field.    The vacuum must remain stable for the reason stated
in section  \ref{freeQFT} above:  it is the only normalizable state
with $p_-=0$.  However, a discussion of  the particle states is more
interesting.
 
In Minkowski space, any massless particle remains stable  because
energy-momentum  conservation forbids its decay.  But energy-momentum
conservation in our background is more subtle.    While each surface
of constant $x^+$ is  homogeneous, the generators of those
translations do not commute with  $\partial_+$.    In particular, the
momenta $p_i = -i \frac{\partial}{\partial  x^i}$ corresponding to
simple translations in $x^i$ do not generate symmetries and are not
conserved.
 
This breaking  of translation invariance should in general   allow any
one-particle state to decay.    This is not particularly remarkable in
the time-dependent case,   as a time dependent background can be
thought of as containing propagating waves with which our particle can
interact.   Let us   therefore concentrate on the time independent
case where $\mu^2_{ij}$ is constant.

To orient ourselves we begin by considering those $\mu^2_{ij}$ which
are positive definite.  Here the momenta $p_\pm$ are positive  and
$p_+$ has a discrete spectrum.  In fact, the mode with lowest $p_+$
has $p_+ >0$.  As a result, a particle in this lowest mode $p_+$ is
again clearly stable by conservation of $p_+$.   This is also true
of strings in the maximally supersymmetric plane wave \cite{BMN}.

In contrast, when $\mu_{ij}^2$ has a negative eigenvalue, the spectrum of
$p_+$ is unbounded below and energy-momentum conservation allows any
particle to decay.  Furthermore, any decay  products will be dispersed
by the repulsive tidal forces removing any hope of stability being
restored by interactions among these products.  It is clear that
an infinite number of particles will be created as $x^+ \rightarrow
\infty$.  As a result, the S-matrix will be non-unitary and
the usual scattering theory  to be inappropriate\footnote{This tendency to
decay infinitely was  briefly discussed in \cite{GPZS}. 
It was also 
noticed that for some string backgrounds the region for which $\mu_{ij}^2$ 
is negative allows a dual description in which $\mu_{ij}^2$ is positive. 
One may take this as evidence that the underlying dynamics is well-defined, 
though typically issues remain associated with objects which are
now non-perturbative in the new dual description.}.   
However, there remains the
possibility of the existence of a  self-adjoint Hamiltonian   which
implements unitary time evolution or some other  well-defined notion
of time evolution over {\it finite} periods of time.
We argue that
this is the case in two ways: first by imposing  an infra-red cut-off,
and second by analogy with a more familiar system  with similar
properties.

\begin{enumerate} 
 
\item  One can impose an infra-red cut-off by placing the system  in a
box.  This makes the spectrum of $p_+$ both bounded below  and
discrete.  As a result, the lowest one-particle states are stable.
Since the infra-red regulated theory is sufficient to describe  the
local dynamics of the full theory within small regions of spacetime,
this local dynamics is well-defined.
 
\item A useful analogy can be constructed by making use of the
observation that our Fock space resembles that of a massless field  in
Minkowski space.    In that case, for any interaction, one-particle
states are prevented  from decay by energy-momentum conservation.  But
suppose that we  break translation invariance by adding a localized
potential (a `scattering center')  near the origin.  The scattering
center can remove excess momentum  so that a massless particle is now
allowed to decay.  Note that massless  particles away from the
scattering center   are stable, so the theory remains unitary  for any
finite number of scattering centers.  However, if we construct an
infinite lattice of scattering centers, then the decay process can
continue ad infinitum and one  expects the theory to  become
non-unitary.  Nonetheless, the theory is  locally equivalent to the
earlier example with a single scattering center.  Thus the model is
well-defined as a local quantum field theory.
\end{enumerate} 
 
In fact, non-unitarity due to infra-red divergences is common place
even in {\it free} quantum field theory on non-compact spaces.  For
example, this feature  arises whenever the background is both time
dependent and spatially homogeneous with non-compact spatial slices.
The time dependence leads  to particle creation, which by translation
invariance must yield some  finite particle number per unit volume.
Since space is infinite in volume,   the total particle number
diverges and the evolution is non-unitary.  This is one reason why
quantum field theory in general curved  backgrounds is often discussed
in terms of algebraic local quantum field theory  (see, e.g.,
\cite{Haag}).  Note that this sort of non-unitary has nothing to do
with  the black hole `information loss paradox'.

One may ask what are the physical effects of the infinite decay of
particles in the plane wave background.  For example, does it lead to
a large back-reaction on the background spacetime?\  In the limit of
weak coupling, the production rate must be slow.    Thus, we expect
that, for weak coupling and if   considered for a short time,  the
particle production instability does not lead to a large
back-reaction.  It seems clear that, although scattering theory   may
not be applicable, there is no reason to regard our spacetime as
leading  to an ill-defined quantum field theory.

\section{String theory in the plane-wave background} 
\label{strings}
 
Since our initial motivation for studying plane waves was their
importance as exactly solvable string backgrounds, it is important to
ask how stringy effects modify the quantum field theoretic picture
described above.  Of particular concern is the fact that the
eigenvalues of $\mu^2_{ij}$ are often described as leading to  mass
terms in the string sigma model, so that one might expect negative
eigenvalues to lead to tachyons and instabilities.  We shall argue
this is not the case, though some  interesting behavior does result.

We begin by briefly  reviewing the quantization of the string in  an
arbitrary time-independent plane wave
background.  We use the Brinkmann  coordinates
$x^\mu$ (so that the spacetime metric again takes the form
\ref{metric}) and choose worldsheet coordinates ($\tau,\sigma$)  with
$\tau$ an (as yet unknown) function of $x^+$, and  $\sigma$ an
orthogonal spacelike coordinate normalized to have period $2\pi$.  The
remaining gauge freedom is fixed   by requiring  the worldsheet metric
to be conformal to the Minkowski metric.  The  function $\tau(x^+)$ is
then determined by the equations of motion.
 
Varying the sigma-model action with respect to $x^-$ leads to  the
equation
 
\begin{equation} 
(-\partial_\tau^2 + \partial_\sigma^2) x^+ =0.
\end{equation} 
Note that the only solutions consistent with the gauge fixing
specified above are  $x^+(\tau) = - \frac{1}{2} p_- \tau + x^+(0)$ for
some constants  $p_-$ and $x^+(0)$.
 
For simplicity, let us assume that the matrix $\mu^2_{ij}$ is diagonal,
with  eigenvalues $\mu^2_i$.   The equation of motion for a transverse
coordinate $x^i$ is then
 
\begin{equation} 
\label{xEOM} 
(-\partial_{\tau}^2 + \partial_{\sigma^-}^2) x^i  -  \frac{1}{4}
p_-^2\mu^2_i x^i
=0,
\end{equation} 
so that the eigenvalue $\mu^2_i$ contributes a squared-mass to  the
field $x^i$.  While $\mu^2_i$ itself can be rescaled  by
rescaling $x^+$ and $x^-$, the factor of $p_-^2$ scales in the
opposite way  so that $p_-^2 \mu^2_i$ is invariant under this
transformation.
 
As usual, it is beneficial to decompose the coordinates $x^i$ into
modes.  We refer to the mode of   $x^i$ with $\sigma$-dependence
$e^{im\sigma}$ as  $x^i_m$.  Taking the Fourier transform of equation
(\ref{xEOM})  yields
 
\begin{equation} 
\partial_\tau^2 x^i_m = (m^2 + \frac{1}{4}p_-^2 \mu^2_i) x^i_m,
\end{equation} 
so that the mode is stable for $m^2 + \frac{1}{4}p_-^2 \mu_i^2 >0$ and unstable
for $m^2 + \frac{1}{4} p_-^2 \mu_i^2 < 0$.  In the latter case, gravitational
tidal forces cause the string to stretch indefinitely in the $x^i$
direction.

Due to the compactness of the string, it is clear that only a finite
number of modes will be unstable for any given values of $p_-$,
$\mu^2_i <0$.    Suppose first that $0> \frac{1}{4}p_-^2 \mu^2_i > - 1$, which,
roughly  speaking, is the case where the spacetime curvature is much
less than  string scale in the reference frame of the
string\footnote{This  interpretation follows by using
(\ref{curv}) and the appropriate Killing field $\xi$ to write   $p_-^2
\mu^2_i = p_-^2 R_{+ i +}^i$   and recalling the relation   $p_- =
\left( \frac{\partial x^+}{\partial \tau}  \right)$.}.  Since only the
zero mode is unstable, the string has no problem retaining its
integrity.  The only instability is that the center of mass of the
string is  rapidly pushed away from $x^i=0$.  One expects that  the
string spectrum will  be discrete, and that different internal
excitations can be interpreted  as particles corresponding to
different quantum fields.  The stability of such fields has already
been discussed in section \ref{FT}.
 
More generally, we may work  out the canonical momenta to find that
the squared masses of string states are given by
 
\begin{equation} 
\label{mass2} 
M^2 =   p_\alpha p_\beta g^{\alpha \beta}  = \sum_{m \ge 0}
\frac{1}{4}p_-^2 [\overline x^i_m \mu_{ij}^2 x^j_m] +  \sum_{m > 0}
\big( \overline p_m^i \delta _{ij}p_m^j + m^2 \overline x^i_m
\delta_{ij}  x^j_m \big),
\end{equation}
where $\overline{x}^i_m$ represents the complex conjugate of $x^i_m.$
Similarly, the dispersion relation takes the form
\begin{equation} 
\label{sdisp} p_+ p_- = \sum_{m \ge 0} \big(\frac{1}{4} 
 p_-^2 [\overline x^i_m \mu_{ij}^2 x^j_m] +  \overline p_m^i \delta
 _{ij}p_m^j + m^2 \overline x^i_m \delta_{ij}  x^j_m \big),
\end{equation} 
One sees that the right hand side of either (\ref{mass2}) or
(\ref{sdisp}) can easily be negative.  However,
this occurs only at large $p_-$.  The dispersion relation  is
qualitatively of the same form as that in section \ref{FT}, where we
saw that it did not in fact lead to instabilities.   Note that
while there is now a new conceptual point associated with
the fact that the locally measured $M^2$ is a function of momentum
(whereas in section \ref{FT} $g^{\alpha \beta}p_\alpha p_\beta$ was
constant), 
this does not in any way effect the analysis of the dispersion relation
(\ref{sdisp}).

As a result, the arguments
given in section \ref{FT} apply here as well.  Said differently, while there are
clearly {\it particle} instabilities in which  the strings are
infinitely stretched or pushed off to infinity, there appear to be no
{\it field theoretic} instabilities in which small field perturbations
grow exponentially in  $x^+$.

However, an interesting phenomenon arises when one investigates the
low energy effective description of strings in this 
background\footnote{In principle, one might wonder if 
the analysis of section \ref{freeQFT} applies to effective fields
with the non-standard kinetic terms associated with the non-standard
dispersion relation (\ref{mass2}).  That no problem arises
is suggested by the similarity of (\ref{sdisp}) and (\ref{disp}), or
may be verified directly in light-cone gauge.  This is particularly
straightforward if one uses a covariant quantization method such as that
of Peierls \cite{Peierls}.}.
One might expect that a low energy limit could be obtained by restricting 
the light cone energy ($p_+$) to be small.  However, since negative
$p_+$ states exist, interactions will generate particles with
arbitrarily large $p_+$ even when the total value of $p_+$ is small.
The same is true if we attempt to restrict to small $p_t$ for
for any $t$ associated
with one of the cut-off `Cauchy surfaces' $\Sigma_\epsilon$
of section \ref{classical}.  
In contrast,  we have seen that $p_-$ is positive for all particles,
so that it makes sense to restrict to small $p_-$.  While only a finite number
of fields can then reach negative $p_+$, the associated low energy theory
nevertheless contains an infinite number of fields.

For completeness, let us briefly discuss the fermionic sector. 
Following \cite{metsaev,tsestring,hong} the fermionic part of the string 
action can be written in terms of the pull back to the 
world sheet of the 
covariant derivative entering in the variation of the gravitino:
\begin{equation} 
D_a =
\partial_a
 + { 1 \over 4}
\partial_a x^m \big[\     (\omega_{ \mu \nu m}
 - {1\over 2}  H_{ \mu  \nu m} \rho_3) \Gamma^{ \mu  \nu}
+
( { 1 \over  3!} F_{\mu\nu\lambda} \Gamma^{\mu\nu\lambda} \rho_1  +
 { 1 \over 2\cdot  5!} F_{\mu\nu\lambda\rho\kappa} 
\Gamma^{ \mu\nu\lambda\rho\kappa}\rho_0 ) \Gamma_m
\  \big]
\end{equation}
where  the $\r_s$-matrices in the $I,J$ space
are the Pauli matrices
$ \rho_1 = \sigma_1$, $\rho_0 = i \sigma_2$.
The action becomes 
\begin{equation}
{\cal L}= i (\eta^{ab} \delta_{IJ} -
 \epsilon^{ab}
(\rho_{3})_{IJ}) \partial_a
  x^{m } \bar \theta^I \Gamma_{m} D_b
\theta^J  
\end{equation}
Here  $\theta^I$ ($I$=1,2)
are the  two real positive chirality  10-d  MW spinors
and $\rho_3=$diag(1,-1). Thus the masses of the 
fermions are completely determined by the value
of the RR form fields. In the simplest case of a background with only
the 5-form field strength non-vanishing $F_5=f(x^+)dx^+\wedge ( \omega_4+
*\omega_4)$, the only nontrivial Einstein equation of motion is
$\rm{Tr}\,\mu_{ij}^2=[f(x^+)]^2$ which does not directly feel the effect of
one negative eigenvalue in $\mu_{ij}^2$. Thus, one may easily deform, 
for example, the BFHP plane wave to one with a negative eigenvalue
without changing the behavior of worldsheet Fermions.  As a result, at
least for appropriate choices of Ramond-Ramond fields, consideration
of Fermions adds nothing new to the picture discussed above.

\section{Discussion} 

In this work we have analyzed the singularity structure of plane wave
metrics and provided a simple criterion to determine the presence of
singularities. In section \ref{sing} we have also presented careful
analysis of the Penrose limit of an AdS background dual to a gauge
theory having a mass scale. Our analysis provides a strictly
gravitational approach to this problem. We have discussed the stability
of classical and quantum
fields in backgrounds having at least one negative eigenvalue of the
matrix $\mu^2_{ij}$. We have found by the introduction of an infrared
regulator that the obvious candidate modes for instabilities are in
fact stable.   While we have given only the most cursory treatment
of boundary conditions, some justification that this is sufficient
was given based on the analyses of plane wave asymptotic 
structures presented in \cite{BN,MR}.

We have also discussed certain features of string theory in plane
waves with negative eigenvalues.  Again, the theory appears
to be stable, though any low energy description necessarily involves
an infinite number of fields.

We have not yet considered the implications of string
interactions.  Let us take a moment to do so now.  Such interactions
trigger an instability of one-particle states in parallel  with that
discussed in section \ref{intQFT}  and the resulting S-matrix will
be non-unitary.  As in the field theoretic case, the issue
might be resolved through a more local treatment.  However, it is an
open question to what extent string theory may be formulated locally
and the difficulties of, for example, using a string field approach to
closed strings are well known.

One difference between the string and field theoretic cases is the
large number of stringy  negative energy states that exist at large
$p_-$.  As we have seen, imposing a strict infra-red cutoff on $x^i$
and an ultra-violet cutoff on $p_-$ discards  most of these states and
truncates the system to having only  a finite number of negative
energy fields.  However, as this cut-off is taken to infinity 
the number of negative
energy fields grows exponentially.  One expects this to have important
effects, at the very least on the thermodynamics of the system.  More
fundamentally, it raises the possibility that non-locality at the
string scale could interact with these negative energy states in a
disastrous  fashion.

On the other hand, it is also possible that string interactions act to
limit the  instability discussed above.  Most of the negative energy
modes  require large $p_-$ but,    since $p_-$ is  positive and
conserved, finite coupling  will induce strings with large $p_-$ to
decay into states with  small $p_-$; i.e., into strings which are more
stable.  It would therefore be interesting  to understand a strong
coupling description of this effect.

There are thus many interesting and fundamental issues raised by
string interactions in negative eigenvalue plane wave backgrounds.  A
proper analysis would seem to require several new techniques, possibly
in a string field theoretic setting.   We hope that the questions
raised by these solvable backgrounds will spur the development of such
techniques.
 
In most of this work we have assumed that at least one eigenvalue
of $\mu^2_{ij}$ is negative and that at least one is positive.
Let us now briefly
address the case where 
all the eigenvalues of the matrix $\mu_{ij}^2$ are negative. 
This case in unphysical if (\ref{metric}) is the Einstein metric
since it violates the weak null energy 
condition. However, it may well arise in the string metric (see, e.g.
\cite{GPZS}) when the dilaton depends on $x^+$.
To analyze the stability of this metric
we note that an explicit coordinate
transformation is known \cite{MR} that maps this space conformally into 
a slice of Minkowski space.  Here, boundary conditions are readily
studied and the results proceed in parallel with those found above.
If desired, this spacetime may again be conformally embedded into the Einstein
static universe, though the resulting map is of very different nature
than the associated conformal embedding of the BFHP plane wave.

\acknowledgments
 
We would like to thank  Daniel Chung, Matthias Gaberdiel, Eric Gimon, Akikazu
Hashimoto, Juan Maldacena,  Horatiu Nastase,  Elias Kiritsis, Boris
Pioline, Massimo Porrati, Jacob Sonnenschein, and Marika Taylor for interesting related
conversations.  We would also especially like to thank Finn Larsen for
motivating us to pursue this project and for his comments on an
earlier draft as well as Dominic Brecher, James Gregory, and Paul
Saffin for sharing with us an early copy of their paper \cite{Dom} and 
for comments on an earlier version of this work.  
D.M. was supported in part by NSF grant PHY00-98747
and by funds from Syracuse University. He would like to thank
the Perimeter institute for their hospitality during part of this work.
L.A.P.Z. is supported by a grant in aid from the Funds for Natural Sciences at
I.A.S.   Both authors  
would also like to thank
the Aspen Center for Physics for their  hospitality during part of
this work.

\end{document}